\documentclass[aps,preprint]{revtex4}%
\usepackage{amsfonts}
\usepackage{amsmath}
\usepackage{amssymb}
\usepackage{graphicx}%
\begin{document}
\title[ ]{Inelastic electron transport through Quantum Dot coupled with an nano
mechancial oscillator in the presence of strong applied magnetic field.}
\preprint{ }
\author{M. Imran$^{\ast}$}
\affiliation{Department of Physics, Quaid-i-Azam University, Islamabad, Pakistan.}
\author{B.Tariq}
\affiliation{Department of Physics, Quaid-i-Azam University, Islamabad, Pakistan. National
Center For Physics, Islamabad, Pakistan.}
\author{M. Tahir}
\affiliation{Department of Physics, University of Sargodha, Sargodha, Pakistan.}
\author{K. Sabeeh}
\affiliation{Department of Physics, Quaid-i-Azam University, Islamabad, Pakistan.}
\pacs{PACS number}

\pacs{PACS number}

\pacs{PACS number}

\pacs{PACS number}

\begin{abstract}
In this study we explain the role of applied magnetic field in inelastic
conduction properties of a Quantum Dot coupled with an oscillator . In the
presence of strong applied magnetic field coulomb blockade effects become weak
due to induced Zeeman splitting in spin degenerate eigen states of Quantum
Dot.By contacting Quantum Dot by identical metallic leads tunneling rates of
spin down and spin up electrons between Quantum Dot and electrodes will be
symmetric. For symmetric tunneling rates of spin down and spin up electrons
onto Quantum Dot, first oscillator get excited by spin down electrons and then
spin up elctrons could excite it further. Where as average energy transferred
to oscillator coupled with Quantum Dot by spin down electrons will further
increase by average energy transferred by spin up electrons to oscillator.
Here we have also discussed that with increasing Quantum Dot and electrodes
coupling strength phononic side band peaks start hiding up, which happens
because with increasing tunneling rates electronic states of \ Quantum Dot
start gettting broadened.

\end{abstract}
\volumeyear{year}
\volumenumber{number}
\issuenumber{number}
\eid{identifier}
\date[Date text]{date}
\received[Received text]{date}

\revised[Revised text]{date}

\accepted[Accepted text]{date}

\published[Published text]{date}

\startpage{1}
\endpage{2}
\maketitle

\subsection{Introduction}

In recent years, much attention has been focused on the concept and
realization of nanoelectromechanical systems (NEMS)\cite{1,2,3,4,5,6,7,8} as a
new generation of quantum electronic devices. A large number of new
experimental techniques have been developed to fabricate and perform
experiments with NEMS in the quantum regime. Examples of high-frequency
mechanical nano-structures that have been produced are nano-scale
resonators\cite{9,10}, semiconductor quantum dots or single
molecules\cite{11,12,13,14,15,16}, cantilevers\cite{18,19}, vibrating crystal
beams\cite{9}, and more recently graphene sheets\cite{20} and carbon
nanotubes\cite{21,22}. These devices are expected to open up a number of
future applications including nanomechanical transport effects, signal
processing which could be used in fundamental research and perhaps even form
the basis for new forms of mechanical computers. Many theoretical methods and
models have been designed in order to account for the behavior of different
types of NEMS system and to make predictions and proposals for future experiments.

In general, there are two different theoretical formulations that can be used
to study the quantum transport in nanoscopic systems under applied bias.
Firstly, a generalized quantum master equation
approach\cite{23,24,25,26,27,28,29,30,31,32} and secondly, the nonequilibrium
Green's function formulation\cite{33,34,35}. The former leads to a simple rate
equation, where the coupling between the dot and the electrodes is considered
as a weak perturbation and the electron- phonon interaction is also considered
very weak. In the latter case one can consider weak and intermediate
electordes to system and electron-phonon coupling. The nonequilibrium Green's
function technique is able to deal with a very broad variety of physical
situations related to quantum transport at molecular levels\cite{36,37}. It
can deal with strong non-equilibrium situations and very small to very large
applied bias. In the early seventies, the nonequilibrium Green's function
approach was applied to mesoscopic transport\cite{38,39,40} by Caroli et al.,
where they were mainly interested in inelastic transport effects in tunneling
through oxide barriers. This approach was formulated in an elegant
way\cite{41,42,43} by Mier et al, where they have shown an exact time
dependent expression for the non-equilibrium current through mesoscopic
systems. In this model an interacting and non-interacting mesoscopic system
was placed between two large semi-infinite leads. In most of the theoretical
work on NEMS devices since the original proposal, the mechanical degree of
freedom has been described classically/semiclassically\cite{23,44} or quantum
mechanically\cite{24,25,26,45,46}using the quantum master or rate equation
approach. In the original proposal, the mechanical part was also treated
classically, including the damped oscillator, and assuming an incoherent
electron tunneling process. This approach is based on a perturbation, weak
coupling and large applied bias approximations, whereas the Keldysh
nonequilibrium Green's function formulation can treat the system leads and
electron-phonon coupling with strong interactions\cite{47}for both small and
large applied bias voltage. The transport properties have been described and
discussed semi-classically/classically but need a complete quantum mechanical
description. A theory beyond these cases is required in order to further
refine experiments to investigate quantum transport properties of NEMS
devices. In the quantum transport properties of these devices; the quantized
current can be determined by the frequency of the quantum mechanical
oscillator, the interplay between the time scales of the electronic and
mechanical degrees of freedom, and the suppression of stochastic tunneling
events due to matching of the Fermionic and oscillator properties.

In the present work, we consider a spin dependent electron transport through a
quantum dot connected to two identical metallic leads via tunneling junctions.
A single nanoelectromechanical oscillator is coupled with quantum dot and gate
voltage is used to tune the levels on the dot. The application of strong
magnetic field induce Zeeman splitting in spin degenrate eigen states of
quantum dot. As a result spin down states moves lower and spin up states move
higher than the degenrate spin eigen states of quantum quantum dot, and thus
offers the different channels of conductance for spin up and spin down
electrons. In the presence of strong applied magnetic field the coulomb
blockade effects will be weak due to Zeeman splitting and we theoratically
included it by mean field approximation, which is quite resonable
approximation for tackling weak interactions. Although electron transport
through mesoscopic systems in the presence of \ Zeeman splitting has been an
active area of research \cite{57, 58, 59} . In our calculation the inclusion
of the oscillator is not perturbative which enable us to predict strong
electron phonons coupling effects in NEMS system. Hence, our work provides an
exact analytical solution to the current-voltage characteristics, conductance,
coupling of leads with the system, and it includes both the right and left
Fermi- level response regimes. However, we have used wide-band
approximation\cite{48,49,50}, where the coupling between leads and quantum dot
is taken to be independent of energy. This provides a way to perform transient
transport calculations from first principles while retaining the essential
physics of the electronic structure of the quantum dot and the leads.

\subsection{Model Hamiltonian}

Our mesoscopic system consists of a Quantum Dot(QD) coupled with an Oscillator
to include the role of phonons effects in conduction through a QD. Application
of external applied magnetic field induce Zeeman splitting in spin degenerate
eigen states of QD. This constitute microscopic part of mesoscopic system.To
incoporate coulomb blockade effects in QD we use mean field approximation
which is useful for weak interaction. As in presence of strong magnetic field
coulomb blockade effects will be small because of Zeeman splitting.
Hamiltonian of the present microscopic system would be,%

\begin{equation}
H_{QD+Oscillator}=\sum_{\sigma}(\epsilon_{\sigma}+\frac{1}{2}\mu_{B}%
g\sigma.B)d_{\sigma}^{\dag}d_{\sigma}+\sum_{\sigma}\alpha_{\sigma}(a+a^{\dag
})d_{\sigma}^{\dag}d_{\sigma}+\omega_{0}(a^{\dag}a+\frac{1}{2}) \label{1}%
\end{equation}

The first term represents two discrete energy levels in QD, which orginates
because of magnetic field induce Zeeman splitting. $d_{\sigma}^{\dag},$
$d_{\sigma}$ $(a^{\dag}a)$ create and annihilate an electron in state
$|\sigma>$ on the dot (create and annihilate a phonon in state $|n>$ on the
oscillator) . Here $\epsilon_{\sigma},$ $\mu_{B},g,\sigma,B,\alpha_{\sigma}$
and $\omega_{0}$ are energy levels of QD electronic-state with spin $\sigma$
,Bohar magneton, Lande g factor,Pauli spin matrix, applied magnetic field, QD
and oscillator coupling and oscillator vibrations frequency. Second term
represents oscillator QD coupling and last term represents oscillator energy spectrum.

In first term of hamiltonian $\left(  \sigma=1\right)  $ for spin up electrons
and $\left(  \sigma=-1\right)  $ for spin down electrons,%

\begin{equation}
\epsilon_{\downarrow}=\epsilon_{o}\ ,\ \ \epsilon_{\uparrow}=\epsilon_{o}+U
\label{2}%
\end{equation}
where $U$ represents coulomb repulsion between spin down and spin up electrons.\ 

To pass current through this sytem we employ left/right electrodes which
constitues macroscopic part of of our system. Hamiltonian of left/right
electrodes is%

\begin{equation}
H_{Leads}=\sum_{k,\sigma}\epsilon_{k,\sigma,\nu}c_{k,\sigma,\nu}^{\dag
}c_{k,\sigma,\nu}. \label{3}%
\end{equation}

Here $\epsilon_{k,\sigma,\nu}$ represents electrodes electronic states with
wave vector $k$, spin $\sigma$ ,and electrodes $\upsilon$ (left/right).
$c_{k,\sigma,\nu\text{ }}^{\dag}(c_{k,\sigma,\nu})$ is electron creation
(annhilation) operator in electrode $\upsilon$ .

Hopping of electrons between electrodes and QD is defined by the following Hamiltonian,%

\begin{equation}
H_{Hopping}=\sum_{k,\sigma}(T_{k,\sigma,\nu}d_{\sigma}^{\dag}c_{k,\sigma,\nu
}+hermitian-congugation) \label{4}%
\end{equation}

Here $T_{k,\sigma,\nu}$ represents electron hopping amplitudes between QD and electrodes.

We first solve our microscopic system Hamiltonian. Our approach include
electron-phonon interaction exactly (non-perturbatively).

To diagonalize microscopic system Hamiltonain , we employ Lang-Firsov
transformation\cite{51}.%

\begin{equation}
\tilde{H}_{QD+Oscillator}=Exp[S]H_{QD+Oscillator}Exp[S^{\dag}] \label{5}%
\end{equation}

where \ \ \ \ \ \
\begin{equation}
S=\sum_{\sigma}(a^{\dag}-a)d_{\sigma}^{\dag}d_{\sigma} \label{6}%
\end{equation}

After diagonalization%

\begin{equation}
\tilde{d}_{\sigma}^{\dag}=d_{\sigma}^{\dag}Exp[\frac{\alpha_{\sigma}}%
{\omega_{0}}(a^{\dag}-a) \label{7}%
\end{equation}

\begin{equation}
\tilde{d}_{\sigma}=d_{\sigma}Exp[-\frac{\alpha_{\sigma}}{\omega_{0}}(a^{\dag
}-a) \label{8}%
\end{equation}

\begin{equation}
\tilde{a}^{\dag}=a^{\dag}-\frac{\alpha_{\sigma}}{\omega_{0}}d_{\sigma}^{\dag
}d_{\sigma}\text{ ,}\tilde{a}=a-\frac{\alpha_{\sigma}}{\omega_{0}}d_{\sigma
}^{\dag}d_{\sigma} \label{9}%
\end{equation}

Therefore,%

\begin{equation}
\tilde{H}_{QD+Oscillator}=\sum_{\sigma}(\epsilon_{\sigma}+\frac{1}{2}\mu
_{B}g\sigma.B-\Delta_{\sigma})d_{\sigma}^{\dag}d_{\sigma}+\omega_{0}(a^{\dag
}a+\frac{1}{2}) \label{10}%
\end{equation}

Where \ \ \ $\Delta_{\sigma}=\frac{\alpha_{\sigma}^{2}}{\omega_{0}}$

Now the eigen function of the diagonalized Hamiltonian in k-space ( eigen
function of harmonic oscillator remain same in real and Fourier's space) would be%

\begin{equation}
|o\sigma n>=\frac{1}{\sqrt{2}}\sum_{\sigma}A_{n}Exp[-\frac{k_{\sigma}^{2}}%
{2}]H_{n}[k_{\sigma}]Exp[-ik_{\sigma}x_{\sigma}]|\sigma> \label{11}%
\end{equation}

\begin{equation}
|u\sigma n>=\frac{1}{\sqrt{2}}\sum_{\sigma}A_{n}Exp[-\frac{k_{\sigma}^{2}}%
{2}]H_{n}[k_{\sigma}]|\sigma> \label{12}%
\end{equation}

$|o\sigma n>$ is state of occupied QD with electron and $|u\sigma n>$ is the
state of un-occupied QD with electron.

Here $x_{\sigma}$ represents displacement of oscillator due to occupancy of
electron in QD. $x_{\sigma}=\frac{\Delta_{\sigma}}{\omega_{0}}$ and
$H_{n}[k_{\sigma}]$ are usual Hermite polynomials.

Now the amplitude of the occupied and un-occupied QD electronic state would be,%

\begin{equation}
A_{mn\sigma}=<o\sigma m|u\sigma n> \label{13}%
\end{equation}

\begin{equation}
A_{mn\sigma}=\frac{1}{\sqrt{\pi2^{m+n+2}n!m!}}%
{\displaystyle\int}
dk_{\sigma}Exp[-k_{\sigma}^{2}]H_{m}[k_{\sigma}]H_{n}[k_{\sigma}%
]Exp[ik_{\sigma}x_{\sigma}] \label{14}%
\end{equation}

\begin{equation}
A_{mn\sigma}=\sqrt{\frac{2^{n-m-2}m!}{n!}}Exp[-\frac{x_{\sigma}^{2}}{4}%
][\frac{ix_{\sigma}}{2}]^{\left\vert n-m\right\vert }L_{m}^{\left\vert
n-m\right\vert }[\frac{x_{\sigma}^{2}}{2}] \label{15}%
\end{equation}

Here $L_{m}^{\left\vert n-m\right\vert }[\frac{x_{\sigma}^{2}}{2}]$ represents
associated Lagurre's polynomials.

After diagonalization tunneling Hamiltonian $H_{Hopping}$ will become,%

\begin{equation}
\hat{H}_{Hopping}=\sum_{k,\sigma}(\hat{T}_{k,\sigma,\nu}d_{\sigma}^{\dag
}c_{k,\sigma,\nu}+hermitian-congugation) \label{16}%
\end{equation}

Where$\left(  \hat{T}_{k,\sigma,\nu}=T_{k,\sigma,\nu}Exp[\frac{\alpha_{\sigma
}}{\omega_{0}}(a^{\dag}-a)]\right)  $

\subsection{Current from the mesoscopic system}

Current from the $\left(  \nu\right)  $ electrode to the QD can be calculated
by taking time derivative of occupation number operator of $\left(
\nu\right)  $ electrode.%

\begin{equation}
J^{\nu}(t)=-e\left\langle \frac{\partial}{\partial t}N^{\nu}\right\rangle
=ie\left\langle \left[  N^{\nu},H\right]  \right\rangle \label{17}%
\end{equation}

where $\left(  N^{\nu}=\sum_{k,\sigma}c_{k,\sigma,\nu}^{\dag}c_{k,\sigma,\nu
}\right)  $ and $\left(  H=\tilde{H}_{QD+Oscillator}+H_{Leads}+\hat
{H}_{Hopping}\right)  $ .Therefore,%

\begin{equation}
J^{\nu}(t)=ie\sum_{k,\sigma}\left(  \hat{T}_{k,\sigma,\nu}\left\langle
c_{k,\sigma,\nu}^{\dag}\tilde{d}_{\sigma}\right\rangle -\hat{T}_{k,\sigma,\nu
}^{\dag}\left\langle \tilde{d}_{\sigma}^{\dag}c_{k,\sigma,\nu}\right\rangle
\right)  \label{18}%
\end{equation}

\begin{equation}
J^{\nu}(t)=i2e\operatorname{Im}[\sum_{k,\sigma}\hat{T}_{k,\sigma,\nu
}\left\langle c_{k,\sigma,\nu}^{\dag}\tilde{d}_{\sigma}\right\rangle ]
\label{19}%
\end{equation}

Now we define electrode and QD coupled lesser Green's function,%

\begin{equation}
G_{un\sigma,k\sigma}^{<}(t,t)=i\left\langle c_{k,\sigma,\nu}^{\dag}d_{\sigma
}\right\rangle \label{20}%
\end{equation}

\begin{equation}
J^{\nu}(t)=i2e\operatorname{Im}[\sum_{k,\sigma}\hat{T}_{k,\sigma,\nu
}G_{un\sigma,k\sigma}^{<}(t,t)] \label{21}%
\end{equation}

To find electrode and QD coupled lesser Green's function we utilize equation
of motion technique (see\cite{43,53,54,55} for utilizing equation of motion
technique in non-equilibrium Green's function theory),%

\begin{equation}
\left(  -i\frac{\partial}{\partial t^{\prime}}-\epsilon_{k,\sigma,\nu}\right)
G_{un\sigma,k\sigma}^{t}(t-t^{\prime})=\hat{T}_{k,\sigma,\nu}^{\dag
}G_{un\sigma,un\sigma}^{t}(t-t^{\prime}) \label{22}%
\end{equation}

where superscript $\left(  t\right)  $ on Green's function notation represents
time ordered Green's function.

Lets define electrode inverse Green's function%

\begin{equation}
g_{k,\sigma,\nu}^{-1}(t^{\prime})=\left(  -i\frac{\partial}{\partial
t^{\prime}}-\epsilon_{k,\sigma,\nu}\right)  \label{23}%
\end{equation}

And by using Green's function identity%

\begin{equation}
g_{k,\sigma,\nu}^{-1}(t^{\prime})g_{k,\sigma,\nu}^{t}(t_{1}-t^{\prime}%
)=\delta(t_{1}-t^{\prime}) \label{24}%
\end{equation}

eq$\left(  \ref{22}\right)  $ can be written in the following form,

\begin{equation}
G_{un\sigma,k\sigma}^{t}(t-t^{\prime})=\hat{T}_{k,\sigma,\nu}^{\dag}\int
dt_{1}G_{un\sigma,un\sigma}^{t}(t-t_{1})g_{k,\sigma,\nu}^{t}(t_{1}-t^{\prime})
\label{25}%
\end{equation}

Now by using analytic continuation rule our electrode QD coupled Green's
function becomes,\cite{56}%

\begin{equation}
G_{un\sigma,k\sigma}^{t}(t-t^{\prime})=\hat{T}_{k,\sigma,\nu}^{\dag}\int
dt_{1}\left[  G_{un\sigma,un\sigma}^{r}(t-t_{1})g_{k,\sigma,\nu}^{<}%
(t_{1}-t^{\prime})+G_{un\sigma,un\sigma}^{<}(t-t_{1})g_{k,\sigma,\nu}%
^{r}(t_{1}-t^{\prime})\right]  \label{26}%
\end{equation}

\bigskip We are discussing dc bias situation therefore its useful to work in
energy space. Hence by using convolution theorem in eq $\left(  \ref{26}%
\right)  $ ,then QD electrodes coupled Green's function will be,%

\begin{equation}
G_{un\sigma,k\sigma}^{t}(\epsilon)=\hat{T}_{k,\sigma,\nu}^{\dag}\left[
G_{un\sigma,un\sigma}^{r}(\epsilon)g_{k,\sigma,\nu}^{<}(\epsilon
)+G_{un\sigma,un\sigma}^{<}(\epsilon)g_{k,\sigma,\nu}^{r}(\epsilon)\right]
\label{27}%
\end{equation}

Hence current from the mesoscopic system is,%

\begin{equation}
J^{\nu}(t)=i2e\operatorname{Im}[\sum_{k,\sigma}\hat{T}_{k,\sigma,\nu}\hat
{T}_{k,\sigma,\nu}^{\dag}\int\frac{d\epsilon}{2\pi}\left[  G_{un\sigma
,un\sigma}^{r}(\epsilon)g_{k,\sigma,\nu}^{<}(\epsilon)+G_{un\sigma,un\sigma
}^{<}(\epsilon)g_{k,\sigma,\nu}^{r}(\epsilon)\right]  ] \label{28}%
\end{equation}

Here $\bigskip\left(  g_{k,\sigma,\nu}^{r,<}\right)  $ $G_{un\sigma,un\sigma
}^{r,<}$ represents (electrodes) QD retarded and lesser Green's
functions.$\ \ \ \bigskip$

From equation of motion method electrodes lesser and retarded Green's function
is given by,%

\begin{equation}
g_{k,\sigma,\nu}^{r}(\epsilon)=\dfrac{1}{(\epsilon-\epsilon_{k\sigma}+i\eta
)}\ \ ;g_{k,\sigma,\nu}^{<}(\epsilon)=i2\pi\delta\left(  \omega-\epsilon
_{k\sigma}\right)  f(\epsilon) \label{29}%
\end{equation}

and%

\begin{equation}
\hat{T}_{k,\sigma,\nu}\hat{T}_{k,\sigma,\nu}^{\dag}=T_{k,\sigma,\nu
}T_{k,\sigma,\nu}^{\dag} \label{30}%
\end{equation}

We employ the wide-band approximation, where the energy density of the
electrodes is taken to be energy independent and $\sum_{k}\rightarrow
\int_{-\infty}^{\infty}d\epsilon_{k\sigma}$%

\begin{equation}
DT_{k,\sigma,\nu}T_{k,\sigma,\nu}^{\dag}\int_{-\infty}^{\infty}d\epsilon
_{k\sigma}g_{k,\sigma,\nu}^{<}(\epsilon)=i2\pi DT_{k,\sigma,\nu}%
T_{k,\sigma,\nu}^{\dag}\int_{-\infty}^{\infty}\delta\left(  \epsilon
-\epsilon_{k\sigma}\right)  d\epsilon_{k\sigma}=i\Gamma_{\sigma}^{\nu
}f_{\sigma}^{\nu}(\epsilon) \label{31}%
\end{equation}

Here $\left(  D\right)  $ represents constant energy density of electrodes.%

\begin{equation}
DT_{k,\sigma,\nu}T_{k,\sigma,\nu}^{\dag}\int_{-\infty}^{\infty}d\epsilon
_{k\sigma}g_{k,\sigma,\nu}^{r}(\epsilon)=DT_{k,\sigma,\nu}T_{k,\sigma,\nu
}^{\dag}\int_{-\infty}^{\infty}\dfrac{d\epsilon_{k\sigma}}{(\epsilon
-\epsilon_{k\sigma}+i\eta)}\ =-i\dfrac{\Gamma_{\sigma}^{\nu}}{2} \label{32}%
\end{equation}

By using eqs $\left(  \ref{29}\right)  $- $\left(  \ref{32}\right)  $in
eq$\left(  \ref{28}\right)  $ our current expression,

\begin{equation}
J^{\nu}(t)=ie[\sum_{\sigma}\int\frac{d\epsilon}{2\pi}\Gamma_{\sigma}^{\nu
}\left[  \left(  G_{un\sigma,un\sigma}^{r}(\epsilon)-G_{un\sigma,un\sigma}%
^{a}(\epsilon)\right)  f_{\sigma}^{\nu}(\epsilon)-G_{un\sigma,un\sigma}%
^{<}(\epsilon)\right]  ] \label{33}%
\end{equation}

For dc transport current will be uniform $J=J^{L}=-J^{R}$ ,So symmetrize
current expression will be $J=\dfrac{J^{L}-J^{R}}{2}$ ,

\begin{equation}
J(\epsilon)=\frac{ie}{2}\sum_{\sigma}\int_{-\infty}^{\infty}\frac{d\epsilon
}{2\pi}[\{\Gamma^{L}(\epsilon)-\Gamma^{R}(\epsilon)\}G^{<}(\epsilon
)+\{G^{r}(\epsilon)-G^{a}(\epsilon)\}\{f_{\sigma}^{L}(\epsilon)\Gamma
^{L}(\epsilon)-f_{\sigma}^{R}(\epsilon)\Gamma^{R}(\epsilon)\} \label{34}%
\end{equation}

Here bold face representations of level width functions $\Gamma^{L/R}$ 's and
Green's function $G^{r,a,<}$shows their matrices in microscopic part of the system.

Now mesoscopic system Green's function could be found from spectral
representation of Green's function,%

\begin{equation}
G_{un\sigma}^{r,a}(\epsilon)=<un\sigma|(\epsilon-\tilde{H}_{QD+Oscillator}I\pm
i\Gamma_{\sigma})|un\sigma> \label{35}%
\end{equation}

where $\Gamma_{\sigma}=\dfrac{\Gamma_{\sigma}^{L}+\Gamma_{\sigma}^{R}}{2}$ and
$I=\sum_{om\sigma}|om\sigma><om\sigma|$%

\begin{equation}
G_{un\sigma}^{r,a}(\epsilon)=\sum_{om\sigma}\frac{<un\sigma|om\sigma
><om\sigma|un\sigma>}{(\epsilon-\epsilon_{\sigma}-\frac{1}{2}\mu_{B}g\sigma
B+\Delta_{\sigma}-(m+\frac{1}{2})\omega_{0}\pm i\Gamma_{\sigma})} \label{36}%
\end{equation}

\begin{equation}
G_{un\sigma}^{r,a}(\epsilon)=\sum_{om\sigma}\frac{|A_{mn\sigma}|^{2}%
}{(\epsilon-\epsilon_{\sigma}-\frac{1}{2}\mu_{B}g\sigma B+\Delta_{\sigma
}-(m+\frac{1}{2})\omega_{0}\pm i\Gamma_{\sigma})} \label{37}%
\end{equation}

While doing numerical calculation we have considered oscillator to be at
ground state so set $\left(  n=0\right)  $, when spin down elctrons comes onto
QD coupled with an oscillator and summation $\left(  m\right)  $ plays the
role of creations of phonons created by spin down electrons on it. While spin
up electrons will come on to QD coupled with an excited oscillator at $\left(
n=m\right)  $ ,excited by spin down electrons. Now summation $\left(
m\right)  $ plays the role of creation of more phonons in oscillator by spin
up electrons.

Mesoscopic system lesser Green's function is given by\cite{47},%

\begin{equation}
G_{un\sigma}^{<}(\epsilon)=iG_{un\sigma}^{r}(\epsilon)(f_{\sigma}^{L}%
(\epsilon)\Gamma_{\sigma}^{L}(\epsilon)+f_{\sigma}^{R}(\epsilon)\Gamma
_{\sigma}^{R}(\epsilon))G_{un\sigma}^{a}(\epsilon) \label{38}%
\end{equation}

Average energy transfer to oscillator by spin up and spin down electrons is
defined by,%

\begin{equation}
E_{PH}=\omega_{0}\left(  \dfrac{\sum_{n}n\rho_{un\sigma}}{\sum_{n}%
\rho_{un\sigma}}\right)  \label{39}%
\end{equation}

In eq$\left(  \ref{39}\right)  $ we have ignored ground state energy
contribution to oscillator, which will give just a shift in average energy
transferred by electrons to oscillator.%

\begin{equation}
\rho_{un\sigma}(t,t)=-i2G_{un\sigma}^{<}(t,t) \label{40}%
\end{equation}

\begin{equation}
G_{un\sigma}^{<}(t,t)=\int\frac{d\epsilon}{2\pi}G_{un\sigma}^{<}(\epsilon)
\label{41}%
\end{equation}

\section{Results And Discussions}

A single level QD in the absence of magnetic field has spin degenerate levels.
As the applied voltage from electrodes become equal to QD energy levels, then
spin up and spin down electrons start tunneling through the QD. Application of
applied magnetic field to QD induce Zeeman splitting. As a result spin up
level moves higher than unsplitted levels and similarly spin down level moves
lower than unsplitted levels. Now as applied voltage from electrodes is
increased first spin down energy level resonates with applied bias and then
spin up energy level resonates with applied bias. In Fig.(1) we have showed
differential conductance as a function of applied bias from electrodes.%

\begin{figure}
[ptb]
\begin{center}
\includegraphics[
height=3.3287in,
width=5.3195in
]%
{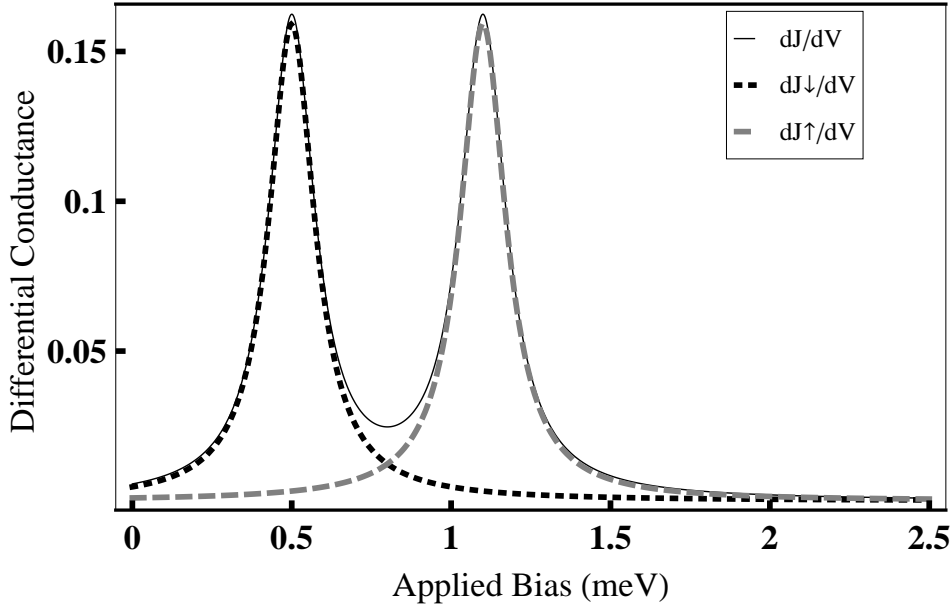}%
\caption{Differential conductance plot with respect to applied bias .Here
$\epsilon_{\downarrow}(B)=0.5$ $meV$ , $\epsilon_{\uparrow}(B)=1$ $meV$ ,
$U=0.1~meV$ .$\Gamma=0.1$ $meV$.}%
\end{center}
\end{figure}

Now we explain QD coupling with an oscillator in the presence of magnetic
field. At zero temperature oscillator will be in ground state.As spin down
electron comes onto QD it gives energy to oscillator and oscillator moves to
excited state. This explain the phononic peaks appearance in differential
conductance for spin down electrons. The main peak of spin down electron will
get shifted due to $\left(  \Delta\right)  $ and amplitude of main peak of
spin down electron becomes smaller as only spin down electron could give
energy to oscillator in zero temperature. We have assumed strong dissipation
effects with enviornment, which means as spin up or spin down electron leaves
oscillator it comes to ground state. This rules out accumulation of energy in
oscillator. When applied bias resonates with $\left(  \epsilon_{\uparrow
}+(m+\frac{1}{2})+U-\Delta\right)  $ then spin up electron channel too get
activated along with spin down electron channel. Here $\left(  m\right)
$represents number of phonons produced by spin down electron, and moreover
spin up and spin down electrons have symmetric coupling with electrodes and QD
$\left(  \Gamma_{\uparrow}=\Gamma_{_{\downarrow}}\right)  $ which means both
spin up and spin down electrons comes onto QD and leaves the QD in the same
time, and this is quite resonable as for identical metallic electrodes QD
electrodes coupling strength will be same for both spin up and spin down
electrons, This could be changed by using ferromagnetic leads\cite{52} .
Therefore spin up electron excites oscillator to even more excited state.So we
get satellite peaks in differential conductance of spin up electrons. Here
phononic peaks of spin up and spin down electrons is not same. This happens as
excited states of occupied QD coupled with oscillator eigen states are not
same for different values of excitation.See Fig(2).

%

\begin{figure}
[ptb]
\begin{center}
\includegraphics[
height=3.3287in,
width=5.3195in
]%
{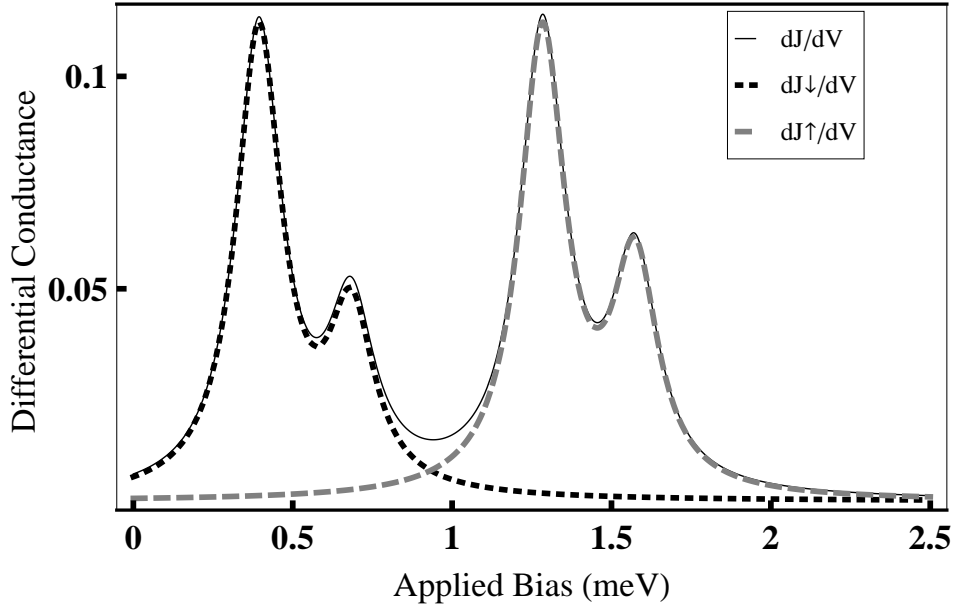}%
\caption{Differential conductance plot with respect to applied bias .Here
$\epsilon_{\downarrow}(B)=0.5$ $meV$ , $\epsilon_{\uparrow}(B)=1$ $meV$ ,
$U=0.1~meV$ ,$\Gamma=0.3$ $meV$,$\alpha=0.37$ $meV$ ,$\omega=0.29$ $meV$.}%
\end{center}
\end{figure}

Effects of \ electrodes-QD coupling $\left(  \Gamma s\right)  $ in the
presence of applied magnetic field and oscillator-QD coupling is of particular
importance. As the tunneling rates of spin up and spin down electrons from
electrodes to QD is increased then energy states of the QD gets broadened. In
Fig.(3) we have showed that for a fixed value of applied \ magnetic field and
QD-oscillator coupling when electrodes-QD coupling are small than Zeeman
splitted peaks and phononic side band peaks are clearly visible. But as we
increase electrodes-QD coupling then phononic side band peaks starts disappearing.%

\begin{figure}
[ptb]
\begin{center}
\includegraphics[
height=3.3287in,
width=5.3195in
]%
{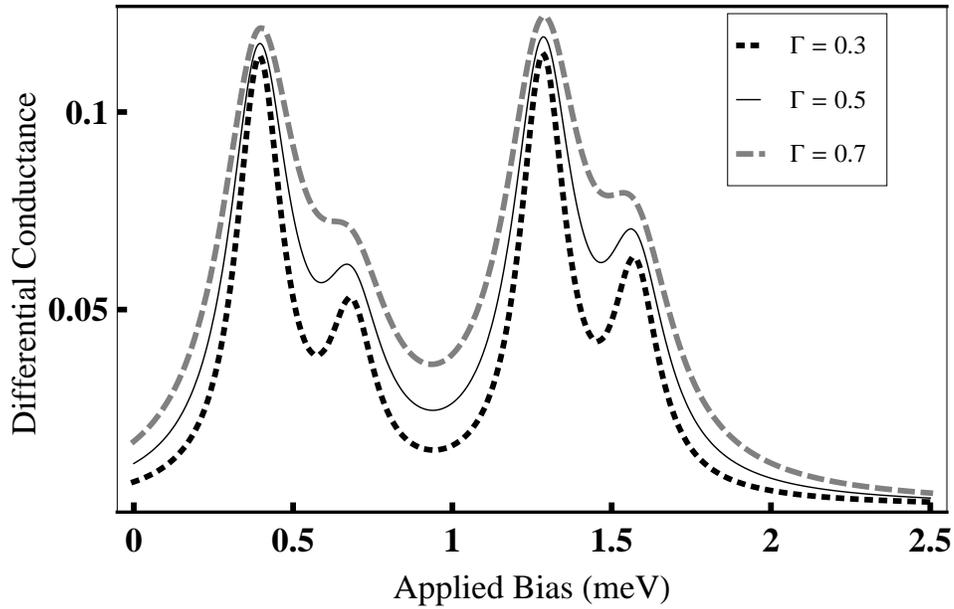}%
\caption{Differential conductance plot with respect to applied bias .Here
$\epsilon_{\downarrow}(B)=0.5$ $meV$ , $\epsilon_{\uparrow}(B)=1$ $meV$ ,
$U=0.1~meV$ ,$\alpha=0.37$ $meV$ ,$\omega=0.29$ $meV$.}%
\end{center}
\end{figure}

Average energy transferred to the oscillator by spin down and spin up
electrons is shown in fig(4). Here we could see that spin down electron curve
lies lower than spin up curve, where as small steps are signature of phonons
creation in averge energy versus applied bias plot. Spin up electron starts
contributing to increase average energy of oscillator where spin down
electrons ends up, and applied bias resonates with $\left(  \epsilon
_{\uparrow}+(m+\frac{1}{2})+U-\Delta\right)  $.%

\begin{figure}
[ptb]
\begin{center}
\includegraphics[
height=2.5028in,
width=3.9885in
]%
{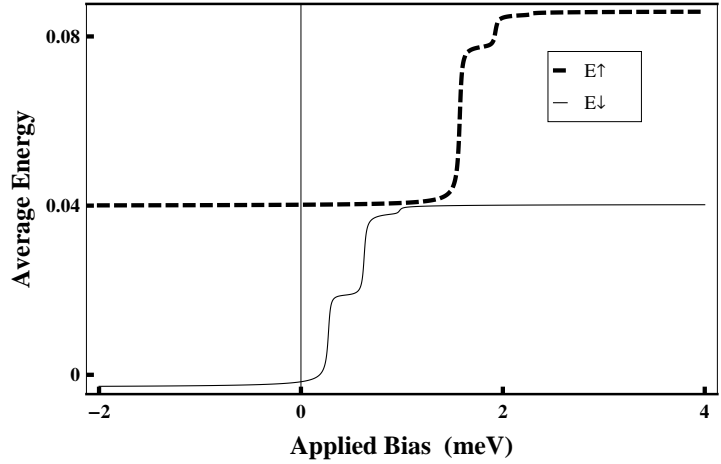}%
\caption{Average energy plot with respect to applied bias .Here $\epsilon
_{\downarrow}(B)=0.5$ $meV$ , $\epsilon_{\uparrow}(B)=1$ $meV$ , $U=0.1~meV$
,$\alpha=0.45$ $meV$ ,$\omega=0.35$ $meV$.}%
\end{center}
\end{figure}

\subsection{Conclusion}

In this work we have studied inelastic electron transport through QD coupled
with an nanomechancial oscillator in the presence of strong applied magnetic
field. We have explained first spin down elctron creates phonon in QD and then
spin up start creating phonons on it. Due to creation of phonons small steps
are produced in average energy transferred to oscillator. With increasing
electrodes QD coupling strength phononic satellite peaks starts hiding up.

\subsection{Acknowledgment}

M. Imran and K. Sabeeh would like to acknowledge the support of the Higher
Education Commission (HEC) of Pakistan through project No. 20-1484/R\&D/09.

$^{^{\ast}}$imran1gee@gmail.com.

\end{document}